\documentclass[10pt,twocolumn,letterpaper]{article}

\usepackage{iccv}

\makeatletter
\@namedef{ver@everyshi.sty}{}
\makeatother

\usepackage{times}
\usepackage{epsfig}
\usepackage{graphicx}
\usepackage{amsmath}
\usepackage{amssymb}
\usepackage{booktabs}
\usepackage{multirow}

\usepackage{tikz}
\usepackage{standalone}
\usepackage{subfig}

\usepackage[accsupp]{axessibility}  
\usepackage{authblk} 

\usepackage[pagebackref=true,breaklinks=true,letterpaper=true,colorlinks,bookmarks=false]{hyperref}

\iccvfinalcopy 

\def\iccvPaperID{26} 

\ificcvfinal\fi

\begin{document}

\title{Self-Supervised Representation Learning using Visual Field Expansion on Digital Pathology}

\author[1]{Joseph Boyd}
\author[1]{Mykola Liashuha}
\author[2]{Eric Deutsch}
\author[3]{Nikos Paragios}
\author[1]{Stergios Christodoulidis\thanks{These authors contributed equally to this work.}}
\author[1]{Maria Vakalopoulou$^*$}

\affil[1]{MICS Laboratory, CentraleSup\'elec, Universit\'e Paris-Saclay,
 91190 Gif-sur-Yvette, France \authorcr{\small\tt firstname.lastename@centralesupelec.fr}}
 

\affil[2]{Department of Radiotherapy, Gustave Roussy Cancer Campus, 94800 Villejuif, France \authorcr
  {\tt\small eric.deutsch@gustaveroussy.fr}}

\affil[3]{Therapanacea, 75014 Paris, France \authorcr
  {\tt\small n.paragios@therapanacea.eu}}

\maketitle
\ificcvfinal\thispagestyle{empty}\fi

\begin{abstract}
The examination of histopathology images is considered to be the gold  standard  for  the  diagnosis  and  stratification  of  cancer patients. A key challenge in the analysis of such images is their size, which can run into the gigapixels and can require tedious screening by clinicians. With the recent advances in computational medicine, automatic tools have been proposed to assist clinicians in their everyday practice. Such tools typically process these large images by slicing them into tiles that can then be encoded and utilized for different clinical models. In this study, we propose a novel generative framework that can learn powerful representations for such tiles by learning to plausibly expand their visual field. In particular, we developed a progressively grown generative model with the objective of visual field expansion. Thus trained, our model learns to generate different tissue types with fine details, while simultaneously learning powerful representations that can be used for different clinical endpoints, all in a self-supervised way. To evaluate the performance of our model, we conducted classification experiments on CAMELYON17 and CRC benchmark datasets, comparing favorably to other self-supervised and pre-trained strategies that are commonly used in digital pathology. Our code is available at \href{https://github.com/jcboyd/cdpath21-gan}{https://github.com/jcboyd/cdpath21-gan}.
\end{abstract}

\section{Introduction}

The characterization and quantification of tissue using microscopy images is considered the gold standard for diagnosis and prognosis in evaluating treatment response for patients with cancer. Traditionally, clinical pathologists examine thin tissue slices under a microscope identifying known biomarkers such as cancer cells, cancer subtypes, percentage of tumour infiltrating lymphocytes, and others. These processes are tedious and time consuming, and moreover may suffer from inter- and intra-observer variability. Currently, with the recent efforts of the computational pathology community, the digitization of tissue slides to whole slide images (WSI) fit for automated analysis is rapidly growing, while more and more research is focused on developing algorithms that can provide accurate and robust tools for clinical use. In particular, with the recent advances in deep learning the automatic analysis of WSIs under supervised and weakly supervised schemes have become very popular ~\cite{courtiol2018classification,campanella2019clinical,lu2021data,van2021deep}.

Although there has been considerable progress in recent years towards the automatic processing of WSIs and its use in clinical practice, there are still a number of lingering challenges. Firstly, the gigapixel size of WSIs makes the development of tailored machine learning techniques challenging. To address this issue, the processing is mainly performed on a tile level while multiple instance learning (MIL) schemes are often developed to predict different clinical endpoints~\cite{zheng2019encoding,lu2021data, wulczyn2021interpretable}. Furthermore, the size of WSIs makes the annotation process by human experts difficult and time-consuming. Most current methods require some sort of supervision either in the form of fully supervisory signals or by utilizing some weakly supervised scheme, making annotated data essential for the development of robust algorithms. Furthermore, with the shift to deep learning the availability of annotations can also impact the generated representations, limiting the reported performances. On these grounds, some recent methods consider the use of pre-trained representations, usually obtained by ImageNet pre-training without investing in generating representations that are specific to WSI images~\cite{gamper2020multiple}.

Unsupervised or self-supervised approaches have recently been studied as an alternative to fully supervised methods, with promising results. Such methods can eliminate the need for annotations and as such can greatly increase the amount of effective training data. With an appropriate problem formulation, self-supervised or unsupervised signals can be leveraged so as to extract compact and informative representations~\cite{sahasrabudhe2020self,vstepec2020image,hou2019sparse}. Traditionally, however, these methods report lower performance than fully supervised ones making their use at present less popular, as they cannot be integrated into clinical practice.

\begin{figure}
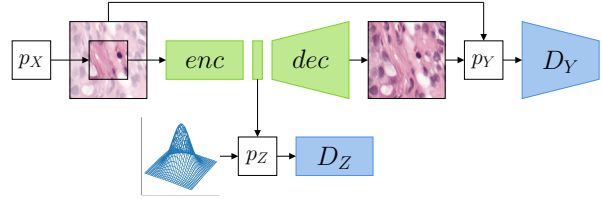

\centering
  \includestandalone[width=0.45\textwidth]{tikz/aae}
  \caption{Learning representations through visual field expansion. An adversarial autoencoder with a progressively grown decoder~\cite{karras2017progressive} and two adversarial losses that ensure a Gaussian distribution for the latent code ($D_Z$) and a realistic visual field expansion ($D_Y$).}
  \label{fig:autoencoder}
\end{figure}

In this study, we propose a novel self-supervised generative method for realistic expansions of the field of view of histopathology image tiles, outpainting the visual context with plausible structure and details (Figure~\ref{fig:autoencoder}). Our method utilizes a progressively grown model that learns robust representations using adversarial training, extrapolating the margins of a tile in a realistic way. In such a setup the network learns in a self-supervised manner the structures and objects that occur in different tissue types within WSIs. The contributions of this work are: \textit{(i)} a novel self-supervised, generative scheme for extrapolating an extended visual field of the tile; \textit{(ii)} a method to generate artificial tiles given specific structures; and \textit{(iii)} generation of histopathology representations that outperform other state-of-the-art pre-trained and self-supervised strategies for histopathology classification tasks. 
To the best of our knowledge, this is the first attempt to explore the use of expansion of the visual field on histopathology slides. Our method outperforms other state-of-the-art and commonly used methods for feature representation in histopathology, reporting performances close to comparable fully supervised methods.

\section{Related work}
Expansion of image borders for semantic image extrapolation is a problem that has previously been investigated by the computer vision community. However, it has not widely been explored by the biomedical imaging community, and in particular within digital histopathology. More specifically,~\cite{yang2019very} proposed an encoder-decoder framework with skip connections and recurrent content transfer for generative natural image scenery prediction. Moreover,~\cite{wang2019wide} proposed a semantic regeneration network based on deep generative models for wide-context semantic image extrapolation on natural images. Our method shares similar ideas with these works, however it is based on progressively grown adversarial autoencoders for the specific task of learning robust feature representations of histopathological images.

In recent years, generative models, and in particular generative adversarial networks (GANs), have been widely used in histopathology~\cite{tschuchnig2020generative}. In~\cite{hou2019robust} the authors proposed a GAN-based method for histopathology image segmentation that synthesizes heterogeneous sets of training image patches of different tissue types. The proposed method enumerates possible ground truth structures during generation of synthetic training patches. In the event the resulting patch is not realistic, the framework decreases its impact in the training loss while, if both realistic and also rarely synthesized, then its impact in the training loss is increased. The method has been validated for the task of nuclei segmentation, proving better generalization compared to other state of the art methods without training data. Recently~\cite{claudio2021pathologygan} proposed the use of GANs to capture key tissue features, structuring these characteristics in its latent space. The authors based their framework on~\cite{jolicoeur2018relativistic,brock2018large,karras2019style} and they show that their model induces an ordered latent space based on tissue characteristics (e.g. cancer cell density or tissue type), allowing to perform linear vector operations
that correspond to high-level tissue tile changes. Contrary to these methods, our framework is based on an encoder-decoder architecture generating high-quality visual content from given structures. Finally, other tasks in which GANs are regularly utilized in histopathology include stain normalization and domain adaptation~\cite{bentaieb2017adversarial,de2018stain}, mainly using the CycleGAN architecture~\cite{zhu2017unpaired}.

The use of self-supervised or pre-trained schemes in histopathology is a common strategy since the amount of annotations is usually limited. Different types of self-supervised problems have been proposed in the literature~\cite{srinidhi2020deep}, focusing on super-resolution and color normalization~\cite{li2021single}, magnification prediction~\cite{sahasrabudhe2020self} as well as contrastive learning~\cite{ciga2020self}. In particular,~\cite{koohbanani2021self} proposed the use of different self-supervised domain-specific auxiliary tasks (Self-Path) such as magnification, jigmad and hematoxylin channel prediction as well as domain-agnostic tasks such as adversarial losses, augmentations and domain prediction. Self-Path reports similar or better performance to supervised baselines on the CAMELYON16 dataset in the case of a low number of annotations. Our work explores another alternative to this direction, learning to predict the invisible while generating robust representations of histopathology images.


\section{Method}

Our model is trained according to the self-supervised task of visual field expansion. Formally, given a target image $\mathbf{x} \in X$ of dimension $w\times h\times C$, we aim to expand the image artificially into a larger image $\mathbf{y} \in Y$ of dimension $W\times H\times C$, where $W > w$ and $H > h$ and such that for some contiguous crop of $\mathbf{y}$, denoted $\mathcal{C}(\mathbf{y})$, we have it that $\mathcal{C}(\mathbf{y}) \approx \mathbf{x}$. In practice, we take the central crop having half the size of the target image in each spatial dimension (thus, $1/4$ of the target image area), and the model is trained to expand from the center outwards. We hypothesise that a model trained for visual field expansion necessitates a rich representation for the observed tissue properties of the input, thereby yielding a powerful encoder of histopathology tiles.

Our proposed model leverages the adversarial autoencoder framework~\cite{makhzani2015adversarial}. An overview of our model is presented in Figure~\ref{fig:autoencoder}. The model consists of an encoder $enc(\cdot) : X \to Z$ mapping image crops $\mathbf{x} \in X$ to a latent vector representation $\mathbf{z} \in Z$, and a decoder $dec(\cdot): Z \to Y$ mapping the latent representation to an expanded image $\mathbf{y}$. The decoder is furthermore a generative model, as the latent representation is trained adversarially against a discriminator network, $D_Z : Z \to \{0, 1\}$ to resemble a known template distribution $p_Z$ . We write,

\begin{multline}
\min_{enc, dec}\max_{D_Z} \mathcal{L}_{AA} = \mathbb{E}_{\mathbf{z}\in p_Z}[\log D_Z(\mathbf{z})] + \\ 
\mathbb{E}_{\mathbf{x}\in p_X}[1 - \log D_Z(enc(\mathbf{x}))] + \\
\mathbb{E}_{\mathbf{y}\in p_Y} [\lambda \cdot R(\mathbf{y}, dec(enc(\mathcal{C}(\mathbf{y}))))]
\end{multline}

where $R(\cdot, \cdot)$ is a reconstruction error function to ensure the autoencoding property and $\lambda$ a manually tuned weight. In practice, we choose $R$ to be the $L_1$ loss, given its aptitude for sharp image synthesis in image-to-image translation tasks~\cite{isola2017image, zhu2017unpaired}. In addition to the convention of alternating generative and discriminative steps, the autoencoder and encoder objectives are trained in alternating reconstruction and regularisation steps. We include an additional discriminator $D_Y : Y \to \{0, 1\}$ on the decoder output images. This proved crucial to ensuring a consistency among the fine details in the original and extrapolated regions. Our full training objective therefore becomes,

\begin{multline}
\min_{enc, dec}\max_{D_Z, D_Y} \mathcal{L} = \mathcal{L}_{AA} + \mathbb{E}_{\mathbf{y}\in p_Y}[\log D_Y(\mathbf{y})] + \\
\mathbb{E}_{\mathbf{x}\in p_X}[1 - \log D_Y(dec(enc(\mathbf{x})))]
\end{multline}

To facilitate the generation of high-resolution images, we adopt the progressive growing algorithm for GANs~\cite{karras2017progressive}. This training algorithm divides model training into stages, beginning with a low resolution target and ``fading in'' targets of increasingly high resolution in each successive stage in a form of curriculum learning. Pioneer networks~\cite{heljakka2018pioneer} are an example of a progressively grown generative autoencoder, however, their training is not GAN-based, and they grow both encoder and decoder symmetrically. In contrast, our model has a fixed encoder taking a fixed $112\times112$ image crop, which we decode progressively.

\section{Experimental setup}

\subsection{Datasets}
In order to evaluate and compare the developed models with the baselines we utilize two publicly available datasets, CAMELYON17~\cite{bandi2018detection} and CRC\footnote{\href{https://doi.org/10.5281/zenodo.1214456}{https://doi.org/10.5281/zenodo.1214456}}~\cite{kather_jakob_nikolas_2018_1214456}. CAMELYON17 consists of $100$ patients, each providing $5$ WSIs. A WSI contains one or more sections extracted from axillary lymph nodes. Of the $500$ slides only $50$ contain metastasis, with the remaining $450$ consisting of negative samples. 

Our pipeline for the CAMELYON17 WSIs follows a standard preprocessing, similar to that described in~\cite{lu2021data}. In practice, the tissue area is first extracted from each WSI in an unsupervised way by converting the WSI to the HSV color space and by using an adaptive Otsu threshold on the saturation channel. Using the resulting masks we crop all tiles of size $224\times224$px in a regular grid within the tissue area at $40$X magnification. From the complete set of tiles, all tiles within metastatic regions were selected and an equal number of normal tiles were randomly sampled. If a tile consisted of more that $50\%$ background pixels it was discarded. We defined positive samples to be a tile that has at least one tumour pixel inside its central $86\times86$px region. A patient-wise splitting was then performed to split the tiles into training, validation and test sets ($70$/$15$/$15\%$ patient-wise). In total, the final training, validation and test sets were respectively $140$k, $30$k and $15$k tiles with equal numbers of positive and negative samples.

The CRC dataset consists of a set of $100$k non-overlapping image tiles from H\&E stained WSIs of human colorectal cancer (CRC) and normal tissue. Train and test splits are provided. We randomly sampled without replacement a validation set amounting to $10\%$ of the training data.  All images are $224\times224$px at $0.5$ microns per pixel (MPP) and are color-normalized using Macenko's method~\cite{macenko}. The tiles are classified into nine tissue type classes: adipose (ADI), background (BACK), debris (DEB), lymphocytes (LYM), mucus (MUC), smooth muscle (MUS), normal colon mucosa (NORM), cancer-associated stroma (STR), colorectal adenocarcinoma epithelium (TUM).

\subsection{Implementation and training details}

\begin{figure*}[t!]%
\centering
\subfloat[Inputs]{\includegraphics[width=0.45\textwidth]{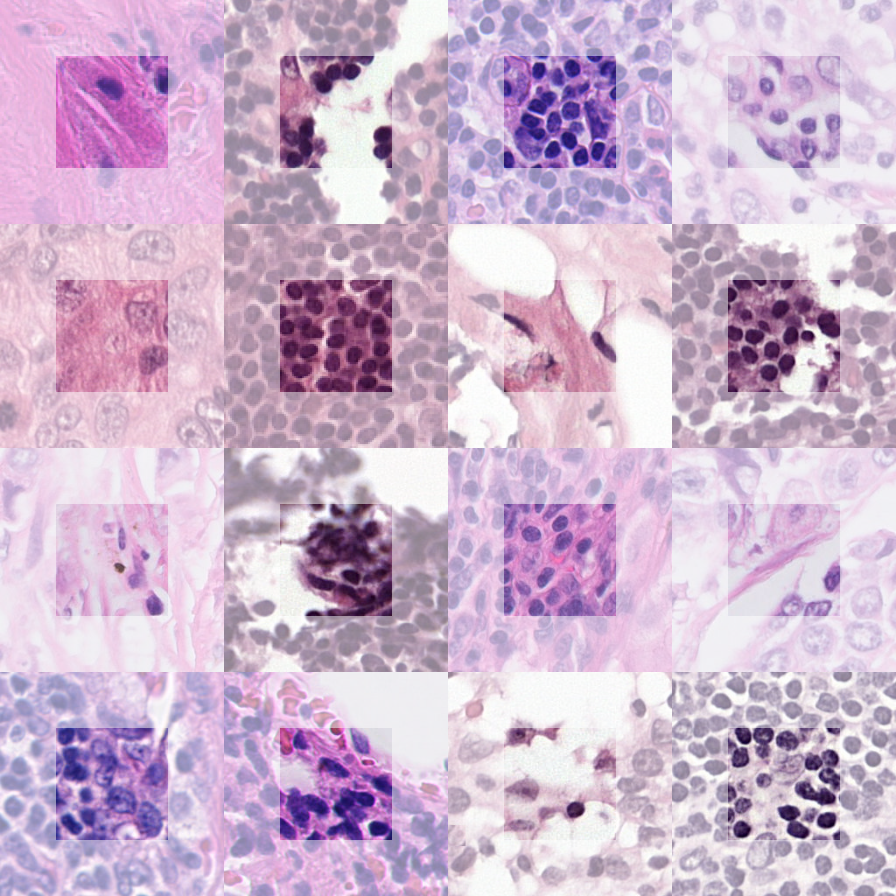}} \qquad
\subfloat[Expansions]{\includegraphics[width=0.45\textwidth]{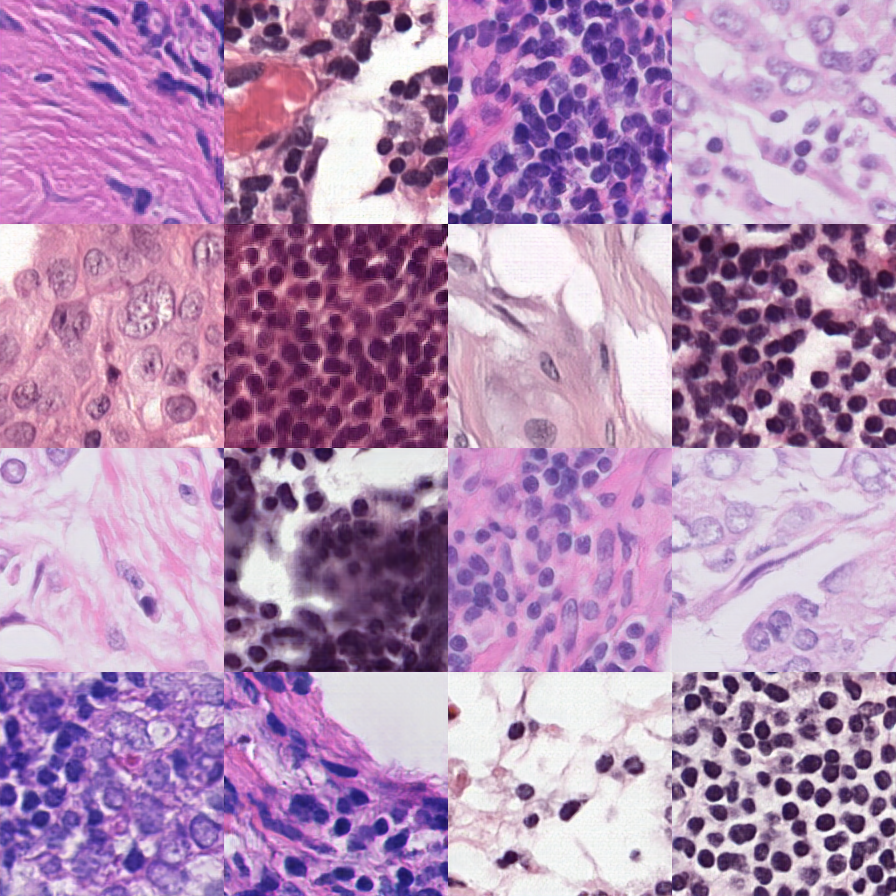}}
\caption{Model inputs (a) and tile expansions (b) from the highest resolution of training ($224\times224$px) for the CAMELYON17 dataset. Only the interior region in (a) are visible to the model.}
\label{fig:reconstructions}
\end{figure*}

To ensure a fair comparison, the encoders of all our models were fixed to be a Resnet18~\cite{he2016deep} with pretrained ImageNet weights. The sole architectural differences between the encoders were the encoder heads, that is, the additional layers appended to each encoder to fulfill the respective training objectives. For our proposed model, this consisted of an additional convolutional and fully-connected projection layer, immediately following the final convolutional layer of the Resnet18 backbone. Given that the generative model is designed to expand the spatial dimensions from input to output, a larger linear ``deprojection'' layer is used in the decoder. Note that the decoder is not used in the downstream feature extraction tasks.

By inspection, we found that freezing the early residual layers of the ResNet18 backbone improved the quality of the model outputs. We manually tuned the freezing strategy to incorporate the first three convolutional blocks, leaving only the final block for fine tuning at training time. Furthermore, to curb the deleterious effects of the large gradients of the randomly initialized model head at the start of training, we lock all backbone weights for the first epoch of training, after which the models are fine-tuned. We used the Adam optimizer ($lr=0.001, \beta_0=0, \beta_1=0.999$) for the training of our proposed method. The progressive training proceeded over five stages of resolution doubling, from $7\times7$px to $224\times 224$px. In the first half of each stage, the new resolution was faded in linearly as a weighted sum of the bilinearly upsampled former and new image resolutions, as in \cite{karras2017progressive}. On average, each stage consisted of the equivalent of approximately $80$ epochs of the training data, with mini-batches of size of $16$.

A computational cluster with NVIDIA Tesla V100 GPUs was utilized for all experiments while PyTorch \cite{NEURIPS2019_9015} and TensorFlow \cite{tensorflow2015-whitepaper} were used for model implementation\footnote{Code available at \href{https://github.com/jcboyd/cdpath21-gan}{https://github.com/jcboyd/cdpath21-gan}} and training. The total training time was approximately $36$ hours.

\section{Results}
\subsection{Evaluation of generated images}
A number of qualitative examples of generated image expansions for CAMELYON17 tiles are presented in Figure~\ref{fig:reconstructions}. On the left we present the inputs to the model together with the ground truth expanded region, and on the right we present the outputs of our model. One may observe that our network is able to realistically generate specific global structures, in addition to fine details, closely approximating the hidden regions. Indeed, the structure and content of the tiles are presented and reproduced accurately. In supplementary materials we separately provide a video of tiles sampled directly from the decoder from template Gaussian noise, over the course of the progressive training.


\begin{table}[t!]
    \centering
    \begin{tabular}{lcc}
    \toprule
         & CAMELYON17 & CRC\\
    \midrule
        FID sampled & 33.37 & 50.47\\
        FID expanded &  20.76 & 37.05\\
    \bottomrule
    \end{tabular}
    \caption{FIDs for generated and expanded tiles for the proposed model on the CAMELYON17 and CRC datasets.}
    \label{tab:fid}
\end{table}

Aside from the qualitative evaluation of the generated images, we calculate the Fr\'{e}chet Inception Distance (FID). FID evaluates the quality of the image outputs of a generative model. In practice, FID compares the feature representations generated by an Inception network for real and generated images. We use the \texttt{pytorch-fid} library \cite{Seitzer2020FID} to evaluate the quality of both generated images and image expansions. In the first case, we generate synthetic images by randomly sampling $512$-dimensional Gaussian noise and decoding it directly into a histopathology tile with the decoder network. As recommended by \cite{heusel2017gans} we compute FID between $50$k generated images and all training images. In the second case, input crops are fed through the full model to create a set of image expansions. These are used in place of sampled images and the FID is computed as before. Table \ref{tab:fid} shows the results of these two evaluations on each of the two datasets. In particular, we note a better performance for expansion than generation. This suggests some degree of divergence in the target and learned distributions of the latent variables. We hypothesise that a small amount of non-Gaussian information, undetected by $D_Z$, is encoded by $enc$ to achieve better reconstructions. We thus characterise a tension between the training objectives of accurate reconstruction and the distributional properties of the latent space. Finally, we observe that our reported FIDs are lower in the case of CAMELYON17 (breast cancer) than the CRC (colorectal cancer) something that is in accordance with the experiments reported in~\cite{claudio2021pathologygan} for the same cancer types.

\subsection{Evaluation of learned representations}
To assess the quality of the learned representations, we used the latent codes $\mathbf{z}$ of our models as features in a pair of downstream classification tasks. For the CAMELYON17 dataset, this was the binary classification of tiles into metastatic and non-metastatic classes; for CRC, this was the classification of tiles into the $9$ tissue types. For each classification task we trained a simple logistic model over training, validation, and test splits of each of the datasets. This model was implemented using the \texttt{scikit-learn}~\cite{pedregosa2011scikit} library setting the maximum iterations to $1500$ to ensure convergence. We report the performance of our model on the CAMELYON17 test dataset using accuracy, precision, recall and F1-score while we report overall balanced accuracy and F1-score for each of the classes of the CRC test dataset.

As baseline feature extractors, we compare our method to another popular self-supervised algorithm, SimCLR \cite{chen2020simple}, as well as a pre-trained ResNet18 network with ImageNet weights. SimCLR minimizes a contrastive loss between encodings of similar patches and maximizes this loss between encodings of different ones. The contrastive loss is computed on mini-batches of paired images, which are strategically augmented in different ways. In this way, one enforces the encoder to find features that better equip the model to distinguish between different patches. For the training of SimCLR, the applied augmentations were color jitter, random rotation and horizontal flipping, color dropping, HED augmentation \cite{heduagm}, image cutout \cite{cutout}, Gaussian noise, Gaussian blur and random cropping. The model was trained using SGD with Nesterov momentum and a batch size of $256$ for $100$ epochs. Two $256$-dimensional fully-connected layers were appended to the $512$-dimensional global pooling of the ResNet18 backbone. Note that these dense layers were discarded after training as in~\cite{chen2020simple}.

\begin{table}[t!]
\begin{adjustbox}{width=0.98\columnwidth,center}

\begin{tabular}{lcccc}
\toprule
     Representation &  Accuracy &  Precision &  Recall &  F1-score \\
\midrule
  Supervised $\dagger$ &     \textbf{90.88\%} &      92.10\% &   \textbf{89.43\%} &     \textbf{90.75\%} \\
         Supervised    &     85.26\% &      \textbf{94.00\%} &   75.33\% &     83.64\% \\
\midrule
\midrule
           ImageNet  &     84.08\% &      84.56\% &   \textbf{83.39\%} &     83.97\% \\
             SimCLR\cite{chen2020simple}   &     84.61\% &      86.69\% &   81.78\% &     84.16\% \\
\midrule
Proposed w/o extrapolation    &     85.40\% &      88.74\% &   81.08\% &     84.74\% \\
  Proposed w/o $D_Z$    &     84.12\% &      \textbf{88.90\%} &   77.96\% &     83.07\% \\
            Proposed    &     \textbf{85.69\%} &      88.32\% &   82.25\% &     \textbf{85.18\%} \\
\bottomrule
\end{tabular}

\end{adjustbox}
\caption{Performance of the different models for the test set of the CAMELYON17 dataset. A logistic regression model was used for all feature representations except Supervised~$\dagger$ for which the performance was calculated using the ResNet18 \cite{he2016deep} softmax output. With bold we highlight the best supervised and self-supervised performances.}
\label{tab:cam}
\end{table}

\begin{table*}[]
    \begin{adjustbox}{width=0.98\textwidth,center}
    \begin{tabular}{lcccccccccccc}
    \toprule
\multirow{2}{*}{Representation} & Balanced & \multicolumn{10}{c}{F1-score} \\
 &  Accuracy & ADI & BACK & DEB & LYM &  MUC & MUS & NORM & STR & TUM & Average \\
\midrule
Supervised $\dagger$ &     85.27\% & 83.94\% & \textbf{99.30\%} & 82.52\% & 94.72\% & \textbf{91.07\%} & 61.53\% & 85.49\% & 65.39\% & 91.93\% &        86.26\% \\
Supervised &     \textbf{86.26\%} & \textbf{85.38\%} & 99.24\% & \textbf{84.97\%} & \textbf{95.11\%} & 89.78\% & \textbf{67.77\%} & \textbf{86.76\%} & \textbf{66.17\%} & \textbf{92.45\%} &        \textbf{87.27\%} \\
\midrule
\midrule
ImageNet &     79.00\% & \textbf{91.58\%} & 99.05\% & 76.84\% & 87.82\% & 87.54\% & 60.50\% & 75.91\% & 42.73\% & 82.66\% &        82.27\% \\
SimCLR \cite{chen2020simple} &     76.29\% & 87.93\% & \textbf{99.76\%} & 65.53\% & 90.23\% & 77.54\% & 59.20\% & 76.81\% & 39.62\% & 84.47\% &        80.03\% \\
\midrule
Proposed w/o extrapolation &     75.67\% & 87.20\% & 98.60\% & 54.83\% & 91.62\% & 89.83\% & 55.49\% & 70.46\% & 43.70\% & 83.64\% &        80.28\% \\
Proposed w/o $D_Z$ &     82.66\% & 88.37\% & 99.59\% & 77.26\% & 88.35\% & 89.97\% & 67.06\% & \textbf{80.05\%} & 57.77\% & 85.57\% &        84.51\% \\
Proposed &     \textbf{85.11\%} & 88.58\% & 98.14\% & \textbf{86.87\%} & \textbf{91.86\%} & \textbf{92.61\%} & \textbf{68.64\%} & 79.89\% & \textbf{61.06\%} & \textbf{88.31\%} &        \textbf{86.30\%} \\
\bottomrule
    \end{tabular}
    \end{adjustbox}
    \caption{Accuracy and F1-score for the CRC dataset. A logistic regression model was used for all feature representations except Supervised~$\dagger$ for which the performance was calculated using the ResNet18 \cite{he2016deep} softmax output. F1-score is reported for the different tissue classes i.e., adipose (ADI), background (BACK), debris (DEB), lymphocytes (LYM), mucus (MUC), smooth muscle (MUS), normal colon mucosa (NORM), cancer-associated stroma (STR), colorectal adenocarcinoma epithelium (TUM). With bold we highlight the best supervised and self-supervised performances.}
\label{tab:crc}
\end{table*}

As an additional baseline we trained completely supervised models using the same ResNet18 backbone network initialised with identical ImageNet pretrained weights. For these experiments, we appended a final dense classification layer with a softmax activation function. Furthermore, during training basic augmentation techniques were applied, such as random flip, random contrast and HED augmentation \cite{heduagm}. All the supervised models were early stopped after approximately 110 epochs. The best model was trained with the AMSGrad variant of the Adam optimizer, categorical cross entropy loss, learning rate of 0.0001, and batch size of 64.

We additionally performed an ablation study on the proposed model. In one variation, we removed the discriminator on the latent code, $D_Z$, allowing the latents $\mathbf{z}$ to assume an arbitrary and unconstrained distribution. In another variation, we replaced the self-supervised image expansion task with simple reconstruction of the input crop. Finally, we trained the model without progressive training, that is, initiating training at the highest resolution. In each case, all other architectural and training properties were identical.

\begin{figure}[t!]
  \centering
  \includegraphics[width=0.45\textwidth]{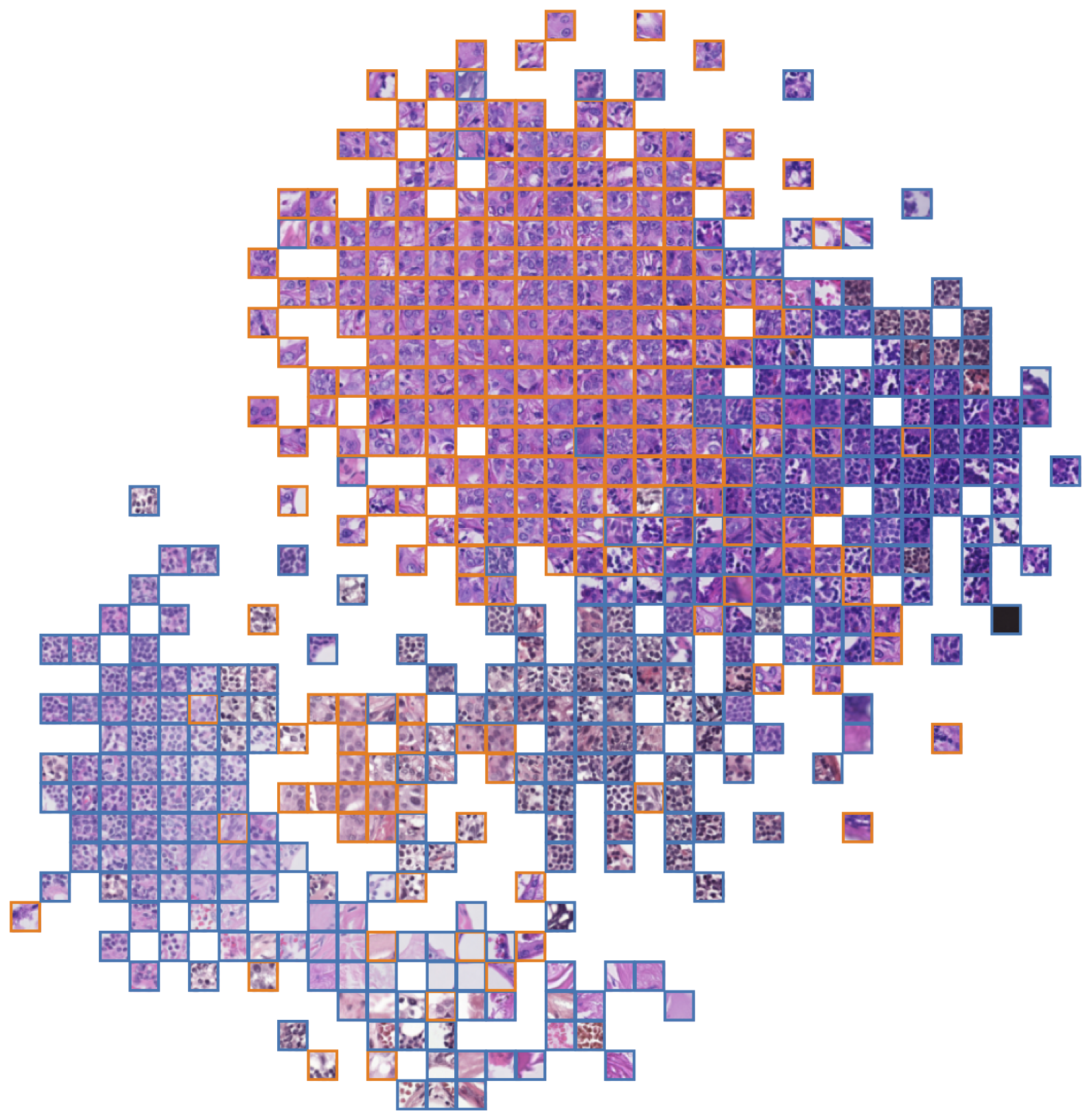}
  \caption{t-SNE plot of sample tiles from the test set used for the CAMELYON17 dataset. Metastatic tiles are outlined in orange, non-metastatic tiles outline in blue.}
  \label{fig:tsne}
\end{figure}

As shown in Table~\ref{tab:cam}, all the evaluated models report balanced overall accuracy above $84\%$ for the CAMELYON17 dataset, with the Supervised~$\dagger$ method exhibiting higher accuracy, recall, and F1-score than all other methods. However, our proposed framework reports similar performances, outperforming both the pretrained ImageNet and SimCLR methods in terms of accuracy (+$1\%$) and F1-score (+$1\%$). Indeed, the highly expressive representations that are generated from our image expansion model are depicted in Figure~\ref{fig:tsne}, in which we plot the $2$D t-SNE embedding produced by the features of the proposed model for the two CAMELYON17 classes. We highlight the ground truth labels, metastatic and non-metastatic, in different colors. It can be observed that the two categories have been successfully separated.


\begin{figure}[t!]
  \centering
  \includegraphics[width=0.45\textwidth]{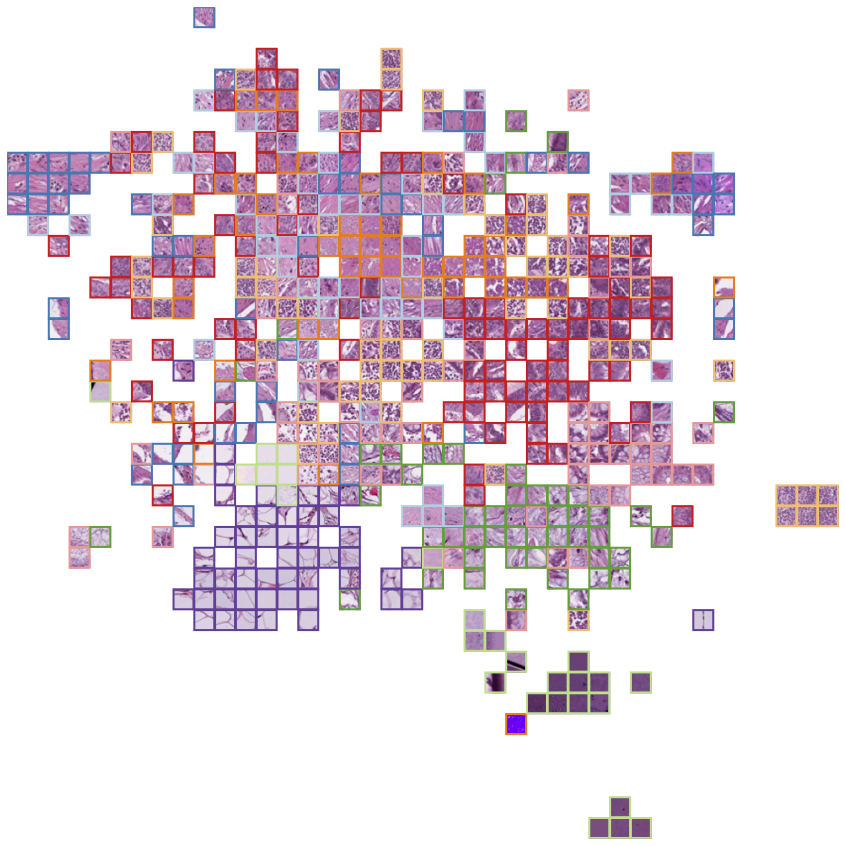}
  \caption{t-SNE plot of samples tiles from the test set used for the CRC dataset. Class is indicated by colour: STR (light blue), MUS (dark blue), BACK (light green), MUC (dark green), NORM (pink), TUM (red), LYM (light orange), DEB (dark orange), ADI (purple)}
  \label{fig:tsne2}
\end{figure}

In Table~\ref{tab:cam} we additionally present the performance of different variations of our method. From these ablation results we observe that both without the discriminator $D_Z$, and separately, without the image expansion task, the model produces less descriptive features for the downstream classification tasks. In the former ablation, we hypothesise that, in imposing a Gaussian prior on the latents, $D_Z$ enacts a regularisation that prevents overfitting to the data. In the latter, we hypothesise that, in the absence of the self-supervised task, the model is less inclined to extract a rich, high-level representation that it can use to generate an image expansion. We forgo the evaluation of the model as trained without progressive training, as in this final ablation, the model tended swiftly towards mode collapse.

The overall accuracy and the performance per class in terms of F1-score for the CRC dataset is presented in Table~\ref{tab:crc}. For these experiments, the supervised methods again outperform the others. However, our proposed method reports a competitive performance, outperforming the rest of the self-supervised methods by more than $6\%$ in terms of overall accuracy. This highlights once more the expressive power of the representations generated by visual field expansion. In Figure~\ref{fig:tsne2} the t-SNE embedding for the CRC dataset is presented. The different classes are again highlighted with different colors, and we can observe a good separation between the different categories. However, in this case we postulate the presence of regions of greater discontinuity in the image manifold. This may follow from the differing sampling policies of the two datasets: the CAMELYON17 tiles were sampled from continuously demarcated regions of tissue, while CRC the tiles are sampled from distinct tissue types. CRC may therefore lend more naturally to a non-Gaussian latent space.

With respect to the different classes that are presented, the smooth muscle (MUS) and cancer-associated stroma (STR) classes report the lowest performance for both supervised and self-supervised methods. Nevertheless, our proposed method reports a similar F1-score to that of the fully-supervised methods. For the remaining classes the F1-score is high for both supervised and self-supervised methods. Overall, in terms of average F1-score, our proposed method performs well and its performance is comparable to the fully supervised one. Once again, the proposed self-supervised formulation with both discriminators outperforms by a significant margin the ablation variants across the majority of classes, highlighting the gain in performance for our proposed design. Conversely, in a minority of classes, the performance of the proposed model is no better than the ablation models. We hypothesise that those classes, in particular background (BACK) and adipose (ADI), contain only sparse tissue, and defy the otherwise rich representations of the self-supervised model to find improvements.



\section{Discussion}

In this paper we proposed a generative model for extending the visual field of histopathology tiles. Our model is grown progressively, while two discriminators ensure a consistency among the fine details in the expanded regions and a structured latent space. The proposed method generates highly realistic images, preserving the structures and content that are presented in specific, predefined input patches. We perform extensive experiments on two publicly available datasets and report a FID of approximately $21$ and $37$ for CAMELYON17 and CRC datasets respectively for the expanded tiles. Moreover, our proposed framework simultaneously learns powerful representations that outperform commonly used pre-trained and self-supervised pipelines on two classification tasks, while reporting comparable performances to equivalent supervised methods. These promising results highlight the effective representation learning of the self-supervised task of visual field expansion.

One limitation of our method is that our framework has been trained such that the visual expansion of the content is performed from the center of the tile. This spatial prior constrains the generated tiles, without fully exploring the potentials of the problem of outpainting. Additionally, we have observed a tradeoff between good reconstruction and generation, with a divergence between the learned latent space and the target distribution. There is therefore scope for future work in reformulating the latent discriminator $D_Z$, or else replacing it with a divergence loss term. On the other hand, we have seen, in particular in the CRC dataset, evidence of non-Gaussianity of the latent space. Adversarial autoencoders are highly flexible models for specifying arbitrary template distributions for the latent space, and are therefore apt to exploring alternative formulations in the future.

\paragraph{Acknowledgements}
This work was supported by the ARC Grant SIGNIT201801286 and LabEx DIGICOSME scholarship (RD N$^o$ 264). Finally we would like to thank Mihir Sahasrabudhe for all the constructive discussions during this project and Mesocenter\footnote{\href{http://mesocentre.centralesupelec.fr/}{http://mesocentre.centralesupelec.fr/}} of CentraleSup\'elec for the computational resources.

{\small
\bibliographystyle{ieee_fullname}
\bibliography{main}
}





\end{document}